\documentclass[prl,twocolumn,superscriptaddress,floatfix,nobalancelastpage]{revtex4}

\usepackage{amsmath}
\usepackage{amssymb}
\usepackage{graphicx}
\usepackage{dcolumn}

\def \basp{BaFe$_2$(As$_{1-x}$P$_x$)$_2$}
\def \bas{BaFe$_2$As$_2$}
\def \bap{BaFe$_2$P$_2$}
\def \sr122{SrFe$_2$P$_2$}

\begin{document}

\title{Enhanced Fermi surface nesting in superconducting BaFe$_2$(As$_{1-x}$P$_x$)$_2$ revealed by de Haas-van Alphen effect}

\author{J.G. Analytis}
\affiliation{Stanford Institute for Materials and Energy Sciences, SLAC National Accelerator Laboratory, 2575 Sand Hill
Road, Menlo Park, CA 94025, USA} 
\affiliation{Geballe Laboratory for Advanced Materials and Department of Applied
Physics, Stanford University, USA}

\author{J.-H. Chu}
\affiliation{Stanford Institute for Materials and Energy Sciences, SLAC National Accelerator Laboratory, 2575 Sand Hill
Road, Menlo Park, CA 94025, USA} 
\affiliation{Geballe Laboratory for Advanced Materials and Department of Applied
Physics, Stanford University, USA}

\author{R.D. McDonald}
\affiliation{Los Alamos National Laboratory, Los Alamos, NM 87545, USA}

\author{S. C. Riggs}
\affiliation{National High Magnetic Field Lab, Tallahassee, FL 87545, USA}

\author{I.R. Fisher}
\affiliation{Stanford Institute for Materials and Energy Sciences, SLAC National Accelerator Laboratory, 2575 Sand Hill
Road, Menlo Park, CA 94025, USA} 
\affiliation{Geballe Laboratory for Advanced Materials and Department of Applied
Physics, Stanford University, USA}

\begin{abstract}
The three-dimensional Fermi surface morphology of superconducting
BaFe$_2$(As$_{0.37}$P$_{0.63}$)$_2$ with $T_c=9K$, is determined using
the de Haas-van Alphen effect (dHvA).  The inner electron pocket has a
similar area and $k_z$ interplane warping to the observed hole pocket,
revealing that the Fermi surfaces are geometrically well nested in the
($\pi,\pi$) direction. These results are in stark contrast to the
Fermiology of the non-superconducting phosphides ($x=1$), and
therefore suggests an important role for nesting in pnictide
superconductivity.
\end{abstract}


\maketitle 

A ubiquitous ingredient in theories of high-temperature
superconductivity is the role of Fermi surface (FS) nesting - a
property originating from the shape of the FS which enhances
quasiparticle scattering along particular directions in momentum
space. FSs that support nesting are typically low dimensional with a
weakly varying morphology in at least one direction in momentum
space. In the case of the pnictides, it is nesting in the ($\pi,\pi$)
direction between electron and hole FSs which is most often invoked
\cite{Chubukov2008a,kuroki_pnictogen_2009,mazin_superconductivity_2009,maier_origin_2009,singh_density_2008}.
From the earliest band structure calculations on the `1111' compounds
e.g. LaFeAsO, it was clear that this was a possible mechanism to
explain both the magnetism and the
superconductivity\cite{singh_density_2008}. However, the `122'
pnictides e.g. \bas\, do not have such a strongly nested FS, yet order
equally strongly in the absence of
doping\cite{mazin_superconductivity_2009}. Furthermore LaFePO, has a
FS which is measured to have well matched electron and hole pockets,
and is therefore close to satisfying a nesting condition, yet has no
known magnetic ordering\cite{coldea,lu}. This has led several authors
to invoke the local magnetic environment, particularly of the As-Fe-As
tetrahedron as a route to understanding these cases
\cite{Yildirim2008a,mazin_superconductivity_2009,
  kuroki_pnictogen_2009}. The role of nesting, is therefore highly
debated.  In order to determine whether a given FS supports nesting,
an intimate knowledge of the full three-dimensional morphology is
required and this is the purpose of the present study.

\begin{figure*}
\includegraphics[width = 15cm ]{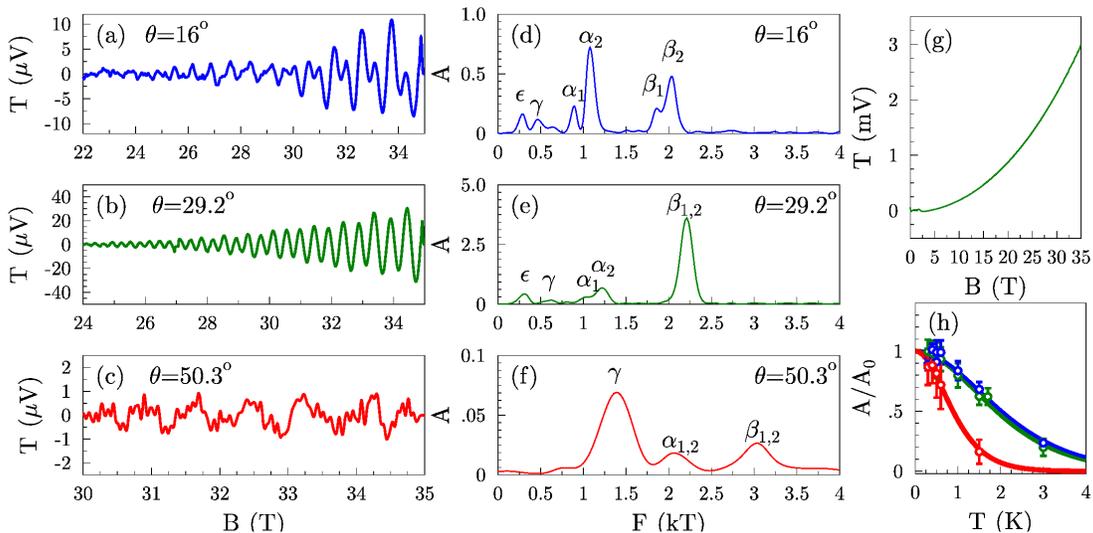}
\caption{The background subtracted raw data (a-c) for three different
  angles taken at $T=0.3$K, and corresponding Fourier spectra in
  (d-f). The Fourier amplitudes of (e) and (f) are normalized with
  respect to (d) for clarity. The $\gamma$ pockets have an enhance
  amplitude near $\theta\approx 51^{\circ}$ probably due to a Yamaji
  effect. (g) gives an example of raw torque data and (h) gives the
  Fourier amplitude as a function of temperature with LK fits for the
  $\beta$ (blue), $\alpha$ (green) and $\gamma$ (red) pockets, from
  which we determine the effective mass. (g) and (h) are taken at
  $\theta=16^\circ$. }
\label{rawdata}
\end{figure*}

We investigate single crystals of superconducting
BaFe$_2$(As$_{0.37}$P$_{0.63}$)$_2$, $T_c=9$K. The compounds \bas\,
and \bap\, are isoelectronic. The former orders antiferromagnetically
at $\sim 138$K, whereas the latter is an ordinary metal down to the
lowest temperatures measured. For intermediate compositions there is a
superconducting dome that peaks at $x\sim 0.3$, with $T_c=30$K. These
materials are particularly useful for dHvA studies because the
disorder is weak and long mean free paths can be maintained. As a
consequence, quantum oscillatory phenomena can be measured, as done
recently by Shishido {\it et al.}  \cite{shishido}. This study made
two important observations: first, the size of the Fermi surface
increases with $x$ beyond optimal As/P substitution and second, the
effective mass decreases. However, due to the resolution limits of the
data the details of the three-dimensional curvature of the Fermi
surface could not be determined for the superconducting compounds and
no hole pockets were observed. In the present paper we resolve the
morphology of the electron pockets and that of one hole pocket. We
conclude that a large amount of the FS is geometrically nested. The
details revealed in this study provide compelling evidence for the
role of nesting as a mechanism for high temperature superconductivity
in this family of compounds.

High quality single crystals of \basp\, with residual in-plane
resistivity ratios $\rho$(300\,K)/$\rho$(1.8\,K) $\sim$20, were grown
from an FeAs flux.  The composition $x=0.63$ was determined by
microprobe analysis. Torque magnetometry was performed using
piezoresistive microcantilevers in fields of up to 35T in DC Bitter
magnets at the NHMFL, Florida. The oscillatory part of the torque
signal ${\rm \mathbf T}\propto {\mathbf M\times \mathbf B}$ originates
from the dHvA effect. Each dHvA frequency is related to an extremal
cross sectional area $A_k$ of the FS in momentum space via the Onsager
relation F=$(\hbar/2\pi e)A_k$. For simply warped, quasi-two
dimensional cylinders two frequencies are expected, a maximum (`belly'
orbit) and a minimum (`neck' orbit). As the angle of the field with
respect to the sample is changed, these extremal orbits traverse
larger sections of the Fermi surface and at particular angles, known
as Yamaji angles, all cross-sectional orbits match, leading to an
enhanced amplitude in the oscillation\cite{yamaji}. Band structure
calculations were performed for the end members \bas\, and \bap\,
using an augmented plane wave plus local orbital method as implemented
in the WIEN2K code \cite{wien2k}. The composition $x=0.63$ does not
have a structural transition at low temperature and therefore the
(room temperature) experimentally determined tetragonal unit cell
parameters are used. These are $a=3.96(3.84)$\,\AA,
$c=13.039(12.44)$\,\AA\, and $z_P=0.3538(0.3456)$ for the As(P)
variants.

Fig.\ \ref{rawdata}(g) shows the the raw torque signal for a single
field sweep up to 35\,T. dHvA oscillations are observed for fields
above $\sim$ 20\,T. In Fig.\ \ref{rawdata} (a-c) we show the
background subtracted data at three typical angles. The Fourier
spectrum is generated by implementing discrete Fourier transform
methods on the signal (in inverse field) for each angle and is shown
in adjacent panels.  At most angles the spectrum is dominated by two
(split) peaks which we label $\alpha_{1,2}$ and $\beta_{1,2}$. In
total five frequencies are observed, which we denote as $\alpha_1,
\alpha_2, \beta_1, \beta_2$ and $\gamma$. We observe a further low
frequency labeled $\epsilon$ which has some scatter about a value of
250T at most angles. However, its value is sensitive to the background
subtraction and the field window used and so it is likely a noise
artifact of the Fourier transform.

\begin{figure}
\includegraphics[width = 8cm]{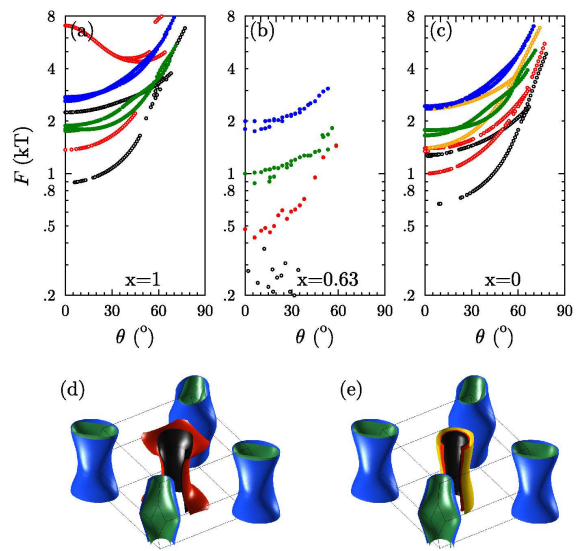}
\caption{(color online). (a) The dHvA angle dependence expected from
  LDA for \bap. (b) the measured angle dependence for a
  superconducting sample of BaFe$_2$(As$_{0.37}$P$_{0.63}$)$_2$. (c)
  The dHvA angle dependence expected from LDA for \bas. The angle
  $\theta=0$ refers to field parallel to the [001] crystallographic
  direction, and $\theta=90$ to the [100]($\theta=90$)
  crystallographic direction. (d) and (e) show the FS as calculated
  for \bap\, and \bas, respectively. The hole pockets are centered at
  the $\Gamma$ point in the zone and the electron pockets at the M
  points.} \label{wien_rot}
\end{figure}

The calculated FS for BaFe$_2$P$_2$ and \bas\, (in the tetragonal
structure) is shown in Figure \ref{wien_rot} (d) and (e)
respectively. Both consist of two similarly warped concentric electron
cylinders at the $M$ point. \bas\, consists of three hole sheets while
\bap\, consists of only two hole pockets centered at the $\Gamma$
point. The hole sheets of \bap\, are much more warped than \bas,
consistent with dHvA measurements on \sr122\cite{analytis_fermi_2009},
suggesting the latter is closer to satisfying a nesting condition. The
dHvA angle dependence associated with these FSs is shown in the Figure
\ref{wien_rot}(a) and (c) for \bap\, and \bas\, respectively.

Figure \ref{wien_rot} (b) shows the angle dependence of each of the
observed frequencies, broadly corresponding to the calculated
band structure of the undoped compounds, though substantial adjustment
is required (see below). It should be noted that this discrepancy
between the experimentally determined Fermi surface cross-sections and
those calculated for the end members ($x = 0\,\&\,1$) can not be
reconciled by using the experimentally determined lattice parameters
for the x = 0.63 compound, which only weakly affect the
calculation. The strongest signal comes from the frequencies labeled
$\alpha$ and $\beta$. In both LaFePO \cite{coldea} and
\sr122\,\cite{analytis_fermi_2009} the strongest signal was from the
electrons pockets, which tend to have the longest mean free paths. It
is natural to assume that this remains the case in the doped compounds
and we assign these frequencies to the electron pockets.

\begin{figure}
\includegraphics[width = 8.2cm ]{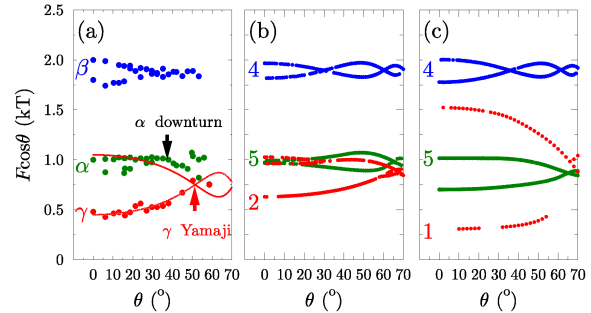}
\caption{(color online) (a) The observed dHvA frequencies plotted as
  $F$cos$\theta$. The observed downturn in the $\alpha$ pocket joining
  the Yamaji angle of the $\gamma$ pocket is shown. (b) The best
  fitting rigid band shifts to the \bas\, band structure with Wien2K
  orbit numbers labeled. The $\beta$ pocket can be broadly matched
  with a shift of -51meV of band 5, the $\alpha$ pocket can be matched
  with a shift of -71meV of band 4 and finally, the $\gamma$ pocket
  can be well approximated with a shift of +25meV to band 2, even
  giving a similar warping to that of the data. (c) Similar shifts to
  the band structure of \bap\, can give good approximation of the
  measured $\beta$ pocket, with a -106meV shift to band 4, $\alpha$
  with a shift of -130meV to band 3. However, though a shift of +84meV
  to band 1 can bring the average frequency into alignment with
  $\gamma$, the pocket is significantly more warped. Larger shifts
  cause band 1 to form a closed 3D FS. In (a) we also show a fit of a
  simple cosine warped cylinder to the $\gamma$ pocket and good
  agreement is achieved with the data, particularly the Yamaji
  angle. } \label{fcostheta}
\end{figure}

The morphology of the pockets is best revealed by plotting the dHvA
frequencies multiplied by cos$\theta$, as shown in Figure
\ref{fcostheta} (a). For a two dimensional FS, the frequency
dependence appears flat, whereas for warped quasi-two-dimensional
cylinders the extremal cross-sections weave between the Yamaji
angles. The $\beta$ pocket appears significantly warped (a difference
in neck and belly frequencies of 175T). $\alpha$ is observable up to
60$^\circ$, suggesting a single pocket exists around 950T. The
assignment $\alpha_1$/$\alpha_2$ is given to the neck/belly of this
pocket and this is confirmed by the comparable effective masses (see
Table \ref{tablemass}). This suggests that the pocket has a weak
warping corresponding to a neck/belly difference of only 85T. The high
angle data is vital for the assignment of the $\gamma$
pocket. Approaching 50$^\circ$ the $\alpha$ frequency abruptly turns
down (see arrow in Figure \ref{fcostheta} (a)), only to appear again
near its average value a still higher angles. Simultaneously, the
$\gamma$ frequency turns upward, joining the $\alpha$ downturn at
around 51$^\circ$. This suggests that for much of the angular range
the $\alpha$ spectrum in fact contains a third frequency, the belly of
the $\gamma$ pocket. Further confirmation of this assignment is
evident in the 51$^\circ$ spectrum where the amplitude of the $\gamma$
is enhanced beyond that of the $\alpha$ frequency, indicating the
first Yamaji angle for this pocket (see Figure \ref{rawdata}
(f)). While it is possible that the $\gamma$ orbit originates from the
same Fermi surface as the $\alpha$ orbit, this seems unlikely given
firstly, that the $\alpha$ orbit remains at the average value of $\sim
950$T at angles $>51^\circ$ and the effective mass of $\gamma$ is
around twice that of $\alpha$ (see below). We conclude that $\gamma$
is a hole pocket. Even without further analysis, it is clear that this
compound has at least one hole pocket which is geometrically well
matched to the inner electron pocket.

As all frequencies deviate substantially from calculations, we perform
rigid band shifts on both the \bas\, and \bap\, Wien2K calculated band
structure, as summarized in Figure \ref{fcostheta} (b) and (c).  The
$\beta$ pocket can be reproduced by the outer electron pocket of
either band structure with relatively moderate shifts in the Fermi
energy (see caption Figure \ref{fcostheta}).  The average frequency of
the $\alpha$ pocket can also be reproduced by the inner electron
pocket of either, but the morphology seems quite different from
observations in both calculations.  The band structure in \bas\, and
\bap\, predicts significantly more warping than is observed, (see
Figure \ref{fcostheta} (b) and (c)).  Similarly fitting the $\gamma$
pocket, only the the \bas\, band structure fits the data adequately,
but again the morphology is off because the Yamaji angle deviates
significantly from the data (by about 15$^\circ$). We finally fit the
$\gamma$ frequency to simple (cosine) warped cylinder. Much better
agreement is achieved with the experimental data and the Yamaji angle
at $\sim 51^\circ$ is almost perfectly matched. In summary, the
experimentally determined FS can support nesting in a manner that is
not obvious from band structure calculations of the end members.

The dHvA mean free path and effective mass are extracted in the
conventional manner using the Lifshitz-Kosevich (LK) formula
\cite{shoenberg} and the results are summarized in Table
\ref{tablemass}. As expected, the hole pocket has a significantly
shorter mean free path than the electron pockets. For the effective
mass, the $\alpha$ and $\beta$ pockets are in good agreement with
those reported in Ref.\onlinecite{shishido}. The effective mass of
$\gamma$ has the largest error bar because of the spectral leakage
from low frequency noise in the Fourier spectrum, which tends to
enhance the mass. Nevertheless, it is clear that the effective mass is
around twice that of the electron pockets. In order to determine the
renormalization we compare our masses to that of the shifted band
structure of \bas\, in Table \ref{tablemass}. We could have equally
chosen to compare to \bap, in which case the renormalization would be
roughly double that shown in the table. The mass enhancements are
different for electron and hole pockets, being about
$\lambda=m^*/m_b-1 \sim 0.8$ for the $\alpha$ and $\beta$ pockets and
$\lambda=m^*/m_b-1 \sim 2.6$ for the $\gamma$ orbit. These
enhancements are much larger than expected from electron-phonon
coupling alone($\lambda_{ep}\simeq 0.25$) \cite{boeri}. Interestingly,
the non-superconducting compounds have electron pockets with a greater
enhancement than the holes, in contrast to the present
case\cite{analytis_fermi_2009}. This study illustrates that as the
superconducting dome is entered this electron/hole mass asymmetry is
reversed, suggesting a vital role for many body effects.

 The total electron count from $\alpha$ and $\beta$ is 0.1 electrons
 per unit cell, whereas the the $\gamma$ pocket accounts for only .03
 holes per unit cell (approximately equivalent to the $\alpha$
 electron pocket), leaving .07 uncompensated holes. As such one
 further hole pocket must exist that is not presently
 observed. Intriguingly, the contribution of the $\alpha$ and $\gamma$
 pocket almost perfectly compensate each other, and so the remaining
 FS must compensate the $\beta$ pocket. This suggests the intriguing
 possibility that the remaining FS is of a similar size to the $\beta$
 pocket, which may therefore sustain FS nesting between the two outer
 electron/hole pockets, in addition to the inner.

\begin{figure}
\includegraphics[width =8cm]{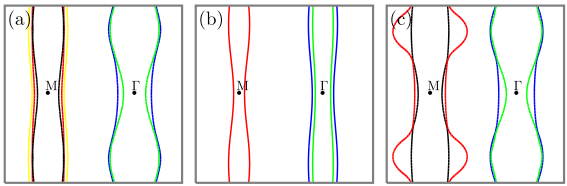}
\caption{(color online).Two dimensional sections through the FS in the
  (110) plane shown for unshifted band structures shown in (a) for
  \bas\, and (c) for \bap. In (b) we show the electron pockets after
  shifting the \bas\, bands to agree with our dHvA, and the a cosine
  warped cylinder is shown for the hole pocket.} \label{slices}
\end{figure}

\begin{table}
\caption{Measured dHvA frequencies, effective masses ($m^*$) and mean
  free paths ($\ell$), along with the values from the band structure
  calculations.  The experimental masses were determined at
  $\theta=16^\circ$.  The band structure values are also estimated at
  the same angle from band structure calculations. Mean-free-paths
  could not be calculated separately for both neck and belly orbits
  due to the data resolution.}
\label{tablemass}
\begin{tabular}{llllllll}
\hline \hline
\multicolumn{4}{c}{Experiment}&\multicolumn{4}{c}{Calculations}\\
 & F(kT) &$\frac{m^*}{m_e}$&$\ell$(nm)& Orbit & F(kT) &$\frac{m_b}{m_e}$ &$\frac{m^*}{m_b}$\\
\hline
$\gamma$    & 0.45      & 4.5(5)       &12 &$2_{\rm min}$ &.53  &1.24& 3.6\\
            &           &               &   &$2_{\rm max}$  &0.86  &1.72&  \\
$\alpha_1$  & .89     & 2.3(1)      &-&$5_{\rm min}$  &1.02  &1.25& 1.8\\
$\alpha_2$  & 1.10     & 2.10(5)      &48&$5_{\rm max}$  &1.04  &1.31& 1.6\\
$\beta_1$   & 1.80      & 2.1(1)     &-&$4_{\rm min}$  &1.81  &1.17& 1.8\\
$\beta_2$   & 2.01      & 2.0(5)     &57&$4_{\rm max}$  &1.95  &1.30& 1.5\\
\hline
\end{tabular}
\end{table}

dHvA studies of the non-superconducting `122' arsenides ($x=0$) reveal
a reconstructed FS\cite{analytisba122}, whereas in the phosphides
($x=1$) they reveal an unresconstructed FS but which is far from
satisfying a nesting condition \cite{analytis_fermi_2009}. In stark
contrast, the present work shows that the intermediate superconducting
compositions have well-matched electron and hole FSs, and are
therefore close to nesting. The differences between our experimental
determination and the band structure calculation for each end member
are summarized in Figure \ref{slices}. Furthermore, the
superconducting compounds have significantly smaller FSs than expected
by band structure, in line with inter-band scattering theory of
Ortenzi {\it et al.} between well-nested Fermi
surfaces\cite{ortenzi_fermi-surface_2009}.  At smaller values of $x$
with a higher $T_c$ it is likely that this trend continues and the
hole pockets even more closely matched the electron pockets, becoming
ever more difficult to observe in dHvA studies and perhaps explaining
their absence in the data of Ref.\onlinecite{shishido}.

In conclusion, we have measured the FS of superconducting
BaFe$_2$(As$_{0.37}$P$_{0.63}$)$_2$ mapping the morphology and mass
enhancement in detail for both electron pockets and one hole
pocket. We find that the warping of the inner electron FS is more
two-dimensional than expected in the unshifted band structure, and
shares a similar degree of warping to that of the hole pocket. The
correlation between favorable nesting and superconductivity presently
observed in \basp\, provides strong evidence for the importance of
nesting in understanding iron-pnictide superconductivity, as suggested
by a number of theoretical
treatments\cite{Chubukov2008a,kuroki_pnictogen_2009,mazin_superconductivity_2009,maier_origin_2009,singh_density_2008}.

The authors would like to thank Antony Carrington for useful comments
and E. A. Yelland for access to computer software. RMcD acknowledges
support from the BES "Science in 100 T" program. Part of this work has
been done with the financial support of EPSRC, Royal Society and EU
6th Framework contract RII3-CT-2004-506239. Work at Stanford was
supported by the U.S. DOE, Office of Basic Energy Sciences under
contract DE-AC02-76SF00515.

%\bibliographystyle{apsrev}
%\bibliographystyle{aps5etal}
%\bibliography{pnictidebib}
\end{document}